\documentclass[12pt]{article}
\usepackage{amssymb,amsmath,epsfig}
\allowdisplaybreaks
\begin{document}

\title{\bf Stability of the Universe Model Coupled with Phantom and Tachyon Fields}
\author{M. Sharif \thanks {msharif.math@pu.edu.pk} and Saadia Mumtaz
\thanks{sadiamumtaz17@gmail.com}\\
Department of Mathematics, University of the Punjab,\\
Quaid-e-Azam Campus, Lahore-54590, Pakistan.}

\date{}
\maketitle

\begin{abstract}
In this paper, we study the phase space analysis of locally
rotationally symmetric Bianchi type I universe model by taking
interactions between dark matter and scalar field models. We define
normalized dimensionless variables to develop an autonomous system
of equations. We also find the corresponding critical points in
order to study the dynamics of the system. The dynamical analysis
indicates that all the critical points correspond to accelerated
cosmic expansion for tachyon coupled field. We observe that positive
values of $m$ provide more stable future attractors as compared to
its negative values. We also analyze the behavior of power-law scale
factor which shows different cosmological phases. It is found that
the region for decelerated expansion gets larger for the phantom
coupled matter by increasing $m$ while this region decreases for
tachyon coupled field.
\end{abstract}

\section{Introduction}

Recent observations suggest that our universe is expanding at an
accelerating rate \cite{1}-\cite{3}. These observational probes
indicate two cosmic phases, i.e., the cosmos phase before radiation
and ultimately the current cosmic era. Many substantial attempts
have been taken to explore the facts behind the current cosmic
acceleration. An unusual form of energy having large negative
pressure, was proposed by the physicists, known as dark energy (DE).
This energy dominates over the matter content undergoing current
cosmic expansion.

There have been a massive literature devoted to study the ambiguous
nature of DE. The cosmological constant ($\Lambda$) is taken as the
elementary candidate having two conventional problems, i.e.,
fine-tuning and cosmic coincidence. Several alternative dynamical
models can be considered as a substitute of $\Lambda$ like
quintessence \cite{4,5}, phantom model \cite{6}, tachyon field
\cite{7} and k-essence \cite{8} that also show the expanding nature
of cosmos. The extension of general barotropic equation of state
(EoS) to some exotic canditates like Chaplygin gas \cite{9} and its
generalization \cite{10} also represent the DE candidates.

The astonishing idea of scalar field, with an EoS parameter other
than $-1$, has been proved to be useful to interpret the universe
evolution due to its applications in different cosmological problems
like cosmic acceleration as well as coincidence problems. The scalar
field candidates also yield the inflationary era of cosmic
evolution. We can adopt scalar field models (in particular, phantom
and tachyon) as an alternative of DE candidates which interact with
DM to resolve the coincidence problem. Researchers have paid an
extensive attention to the tachyon cosmology in which the tachyon is
fundamentally attributed by string theory \cite{11,12}. Gibbons
studied the cosmological impact of tachyon field turning down to its
ground phase \cite{13}. The tachyon matter may execute inflationary
era and eventually a new form of DM at late times \cite{14}. A
phantom field was also presented as an alternative of DE that
constitutes large negative pressure with EoS parameter $w<-1$ and
plays an important role for accelerated expansion of the universe
\cite{15}-\cite{18}. One of the significant features of the phantom
model reveals that the universe will end with a big-rip (future
singularity). The interaction of DE (phantom or tachyon) and DM may
alleviate the coincidence problem \cite{19,20}.

A phase space represents various states including position as well
as momentum corresponding to each point of a dynamical system. This
analysis yields dynamical characterization of a cosmological model
by reducing complexity of the system of equations. It has always
been helpful to study different stages of evolution by converting
the system of equations to an autonomous one. This investigates the
influence of initial data on the stability of a system by checking
whether the respective system will remain stable or not \cite{21}.

Copeland et al. \cite{22} discussed phase space analysis of the
inflationary model that was unable to solve density problem. Guo et
al. \cite{23} studied stability of FRW universe model filled with
barotropic fluid and phantom scalar field. They found that phantom
dominated solution is a stable late-time attractor. Guo et al.
\cite{24} used this analysis to study the dynamical behavior of
cosmos by interacting phantom energy with DM. Chen et al. \cite{24a}
discussed a detailed phase space analysis of various phantom
cosmological models for which the DE sector interacts with the DM.
Acquaviva and Beesham \cite{25} explored this analysis for FRW model
and found that the nonlinear extension of viscous cosmic model
provides accelerated expansion of the universe. We studied the
influence of nonlinear electrodynamics on stability of accelerated
expansion in the background of FRW cosmology \cite{26}. Shahalam et
al. \cite{27} presented phase space analysis for various
combinations of the coupling between phantom and tachyon fields
taking FRW universe. We also discussed phase space analysis of FRW
universe model by considering a power-law model for bulk viscosity
coefficient \cite{28}.

Bianchi universe models have also been discussed in literature to
explore primordial anisotropy and some large angle anomalies
identified by CMBR which give rise to the negligence of statistical
cosmic isotropy \cite{29}. Belinkskii and Khalatnikov \cite{30}
applied phase plane approach to Bianchi type I (BI) model and
studied the effects of shear and bulk viscosity. Coley and Dunn
\cite{31} used this technique to investigate the dynamics of Bianchi
type V model by taking a viscous fluid. Burd and Coley \cite{32}
discussed the influence of shear as well as bulk viscosity on
stability of Bianchi universe models. Goliath and Ellis \cite{33}
studied dynamical evolution of Bianchi universe model by including
the cosmological constant. Sharif and his collaborators discussed
phase space analysis of Bianchi I universe in Brans-Dicke gravity
with chameleon scalar field \cite{33a} as well as $f(T)$ gravity
taking non phantom, vacuum and phantom phases \cite{33b}. Chaubey
and Raushan \cite{34} studied phase space analysis of LRS BI model
in the presence of scalar field.

This paper aims to investigate the stability of LRS BI universe
model by taking linear interactions of phantom and tachyon fields
coupled with DM through phase space analysis. The plan of the paper
is as follows. In section \textbf{2}, we provide some basic
formalism for the evolution equations. An autonomous system of
equations is developed by defining normalized dimensionless
variables. Section \textbf{3} deals with the dynamical analysis of
the universe model by taking three different forms of interactions
between phantom energy and DM. We also study the phase space
analysis of tachyon field coupled with DM in section \textbf{4}.
Section \textbf{5} deals with the formulation of power-law scale
factor. Finally, we summarize our findings in the last section.

\section{General Equations}

Bianchi universe models are the simplest extensions of FRW universe
with the addition of anisotropic effects. The LRS BI model is
described by the line element \cite{35,36}
\begin{equation}\label{1}
ds^2=-dt^2+a(t)dx^2+b(t)(dy^2+dz^2),
\end{equation}
where $a(t)$ and $b(t)$ are the cosmic scale factors. The mean
Hubble parameter, in terms of the directional Hubble parameters
($H_{1}=\frac{\dot{a}}{a}$ and $H_{2}=\frac{\dot{b}}{b}$), can be
defined as
\begin{equation}\label{1a}
H=\frac{1}{3}[H_{1}+H_{2}]=\frac{1}{3}\left[\frac{\dot{a}}{a}
+\frac{2\dot{b}}{b}\right]=\frac{1}{3}\left(\frac{\dot{v}}{v}\right).
\end{equation}
We consider a power-law relation $a=b^m$, $m\neq0,1$, where $m$
represents a constant anisotropic parameter that gives the deviation
of anisotropy from the isotropic model. This leads to the following
relation
\begin{equation}\label{1b}
H_{1}=mH_{2}=\left(\frac{3m}{m+2}\right)H.
\end{equation}
This assumption can be rationalized by the velocity redshift
expression for extragalactic sources through which the Hubble cosmic
expansion may achieve isotropy when shear to expansion scalar ratio
is constant \cite{37}.

We assume a cosmic fluid by interacting phantom field with matter
such that the conservation of energy yields
\begin{eqnarray}\label{4}
\dot{\sigma}_m+3(\sigma_m+p_m)H&=&Q,\\\label{4a}
\dot{\sigma}_{\phi}+3(\sigma_{\phi}+p_{\phi})H&=&-Q,\\\label{4b}
\dot{\sigma}+3(\sigma+p)H&=&0,
\end{eqnarray}
where $Q$ is the interaction term, dot corresponds to the derivative
with respect to time, $\sigma=\sigma_m+\sigma_\phi$, $p=p_m+p_\phi$,
$\sigma_m$, $\sigma_\phi$, $p_m$ and $p_\phi$ represent energy
densities and pressures of the matter and phantom energy,
respectively. It should be noted that the transfer of energy between
two components depends on the sign of interaction term. The energy
passes from phantom to matter for $Q>0$ while $Q<0$ yields vice
versa. The conservation equations clearly show that
$Q=Q(H,\sigma_m,\sigma_\phi)$. The constraint and Raychaudhuri
equations are formulated from the field equations as
\begin{eqnarray}\label{2}
H^2&=&\frac{(m+2)^2}{9(2m+1)}(\sigma_m+\sigma_\phi),
\\\label{3}
0&=&\left(\frac{6}{m+2}\right)\dot{H}+\frac{27}{(m+2)^2}H^2+p_\phi,
\end{eqnarray}
where $\sigma_\phi=-\frac{1}{2}\dot{\phi}^2+V(\phi)$,
$p_\phi=-\frac{1}{2}\dot{\phi}^2-V(\phi)$. Due to many arbitrary
parameters, it seems difficult to find analytical solution of the
evolution equation. For this purpose, we define the following
normalized dimensionless quantities
\begin{equation}\label{5}
\mu=\frac{(m+2)\dot{\phi}}{\sqrt{6}(2m+1)H},\quad
\nu=\frac{(m+2)\sqrt{V}}{\sqrt{3}(2m+1)H}, \quad
\lambda=-\frac{V'}{V},
\end{equation}
that transform the dynamical equations into an autonomous system.
Differentiating $\mu$ and $\nu$ with respect to
$N=\frac{m+2}{3m}\ln{a}$, we have
\begin{eqnarray}\label{6}
\mu'&=&\frac{3m}{m+2}\mu\left[\frac{\ddot{\phi}}{H\dot{\phi}}
-\frac{\dot{H}}{H^2}\right],\\\label{6a}
\nu'&=&-\frac{3m}{m+2}\nu\left[\sqrt{\frac{3}{2}}
\frac{2m+1}{m+2}\mu\lambda+\frac{\dot{H}}{H^2}\right].
\end{eqnarray}
The Raychaudhuri and conservation equations through these
dimensionless parameters become
\begin{eqnarray}\label{7}
\frac{\dot{H}}{H^2}&=&\frac{1}{2(m+2)}\left[-9
+(2m+1)^2\{\mu^2+\nu^2\}\right],\\\label{7a}
\frac{\ddot{\phi}}{H\dot{\phi}}&=&-3-\sqrt{\frac{3}{2}}
\frac{2m+1}{m+2}\frac{\lambda\nu^2}{\mu} +\frac{Q}{H\dot{\phi}^2},
\end{eqnarray}
where $\lambda$ is a constant. The constraint equation can be
written in the form
\begin{equation}\label{8}
\Omega_{\phi}=\frac{(m+2)^2}{9(2m+1)}\frac{\sigma_\phi}{H^2}
=\frac{2m+1}{3}[-\mu^2+\nu^2].
\end{equation}
The effective EoS for the cosmic fluid and phantom field are given
by
\begin{eqnarray}\label{8a}
w_{eff}=-1-\frac{2\dot{H}}{3H^2},\quad
w_{\phi}=\frac{w_eff}{\Omega_{\phi}}.
\end{eqnarray}

\section{Dynamics of Interacting Phantom Energy}

In this section, we discuss dynamics of LRS BI universe by the phase
space analysis taking an interaction between phantom energy and
matter. In this context, we impose the condition $\mu'=\nu'=0$ to
solve the system of Eqs.(\ref{6}) and (\ref{6a}) and also determine
the corresponding critical points $\{\mu,\nu\}$. The stability of
universe model depends on the behavior of critical points and their
eigenvalues. We take three different types of coupling between
phantom field and matter as follows.

\subsection{Coupling $Q=\alpha\dot{\sigma}_m$}

Firstly, we consider $Q=\alpha\dot{\sigma}_m$ as an interaction
model for the universe where both phantom field and DM are coupled.
Equation (\ref{7a}) for this coupling takes the form
\begin{eqnarray}\label{9}
\frac{\ddot{\phi}}{H\dot{\phi}}=-3-\sqrt{\frac{3}{2}}
\frac{2m+1}{m+2}\frac{\lambda\nu^2}{\mu}-\frac{9\alpha
\Omega_m}{2(2m+1)(1-\alpha)\mu^2},
\end{eqnarray}
where $\Omega_m=1-\Omega_{\phi}$. The autonomous system of equations
becomes
\begin{eqnarray}\nonumber
\mu'&=&\frac{3m}{m+2}\mu\left[-3-\sqrt{\frac{3}{2}}
\frac{2m+1}{m+2}\frac{\lambda\nu^2}{\mu}-\frac{9\alpha
\Omega_m}{2(2m+1)(1-\alpha)\mu^2}-\frac{1}{2(m+2)}\right.\\\label{9a}
&\times&\left.\{-9+(2m+1)^2(\mu^2+\nu^2)\}\right],\\\nonumber
\nu'&=&-\frac{3m}{m+2}\nu\left[\sqrt{\frac{3}{2}}
\left(\frac{2m+1}{m+2}\right)\mu\lambda+\frac{1}{2(m+2)}
\{-9+(2m+1)^2(\mu^2+\nu^2)\}\right].\\\label{9b}
\end{eqnarray}
We can find the eigenvalues by the Jacobian matrix
\begin{equation}
A=\left(
\begin{array}{cc}
\frac{\partial f}{\partial \mu} & \frac{\partial f}{\partial \nu}\\
\frac{\partial g}{\partial \mu} & \frac{\partial g}{\partial \nu}\\
\end{array}
\right)_{0},
\end{equation}
where suffix $0$ defines the values at critical points $(\mu_{c},
\nu_{c})$. The fixed point is called a source or unstable point
(respectively, a sink or stable point) if both eigenvalues consist
of positive (respectively negative) real parts. The real parts of
the eigenvalues having opposite signs correspond to a saddle point
of the system. We evaluate the following critical points in this
case. For $P_{1}=(\mu_c,
\nu_c)=\left(\frac{-\sqrt{6}\lambda+\sqrt{6(\lambda^2+6)}}{2(2m+1)},
0\right)$, the eigenvalues of Jacobian matrix are obtained as
\begin{eqnarray}\nonumber
\eta_{1}&=&\frac{9m}{m+2}\left[-1-\frac{3\alpha}{2(1-\alpha)(2m+1)}
\left\{\frac{-4(2m+1)^2}{[-\sqrt{6}\lambda+\sqrt{6(\lambda^2+6)}]^2}
+\frac{2m+1}{3}\right\}\right.\\\label{10}&+&\left.
\frac{3}{2(m+2)}-\frac{[-\sqrt{6}\lambda+\sqrt{6(\lambda^2+6)}]^2}{8(m+2)}\right],
\\\nonumber
\eta_2&=&-\frac{3m}{m+2}\left[\sqrt{\frac{3}{2}}\lambda
\frac{\sqrt{6(\lambda^2+6)}-\sqrt{6}\lambda}{m+2}+\frac{1}{2(m+2)}
\left\{-9\right.\right.\\\label{10a}&+&\left.\left.
\frac{[\sqrt{6(\lambda^2+6)}-\sqrt{6}\lambda]^2}{4}\right\}\right].
\end{eqnarray}
We are interested to explore the influence of physical parameters
($\alpha$ and $m$) on the stability of cosmos coupled with scalar
field. We plot the dynamical characteristics of critical points for
$Q=\alpha\dot{\sigma}_m$ in the vicinity of different values of
$\alpha$ and $m$ as shown in Figure \textbf{1}. We find negative
eigenvalues which show the point $P_{1}$ as a stable future
attractor for $m>0$ in the physical phase space except for
$\alpha=1$ at which the system becomes undetermined. For
$\alpha,~m<0$, this point behaves as an unstable node.

The effective potential for the cosmic fluid is defined as
\begin{equation}\label{16a}
w_{eff}=-1+\frac{1}{m+2}[-9+(2m+1)^2(\mu^2+\nu^2)].
\end{equation}We also evaluate the effective EoS parameter and deceleration
parameter as
\begin{eqnarray}\label{16b}
w_{\phi}&=&\frac{1}{\mu^2+\nu^2}\left[-1+\frac{1}{m+2}\{-9+(2m+1)^2
(\mu^2+\nu^2)\}\right],\\\label{16c}
q&=&-1+\frac{1}{m+2}[-9+(2m+1)^2(\mu^2+\nu^2)].
\end{eqnarray}
The dynamical analysis shows accelerated expansion of the universe
model for $P_{1}$ as $q<0$.
\begin{figure}\center
\epsfig{file=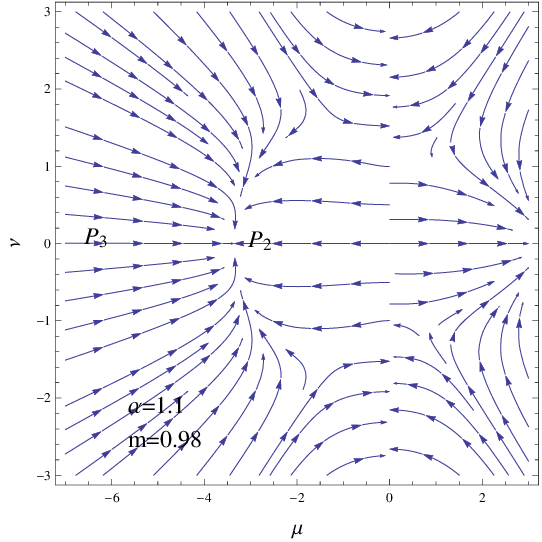,width=0.45\linewidth}\epsfig{file=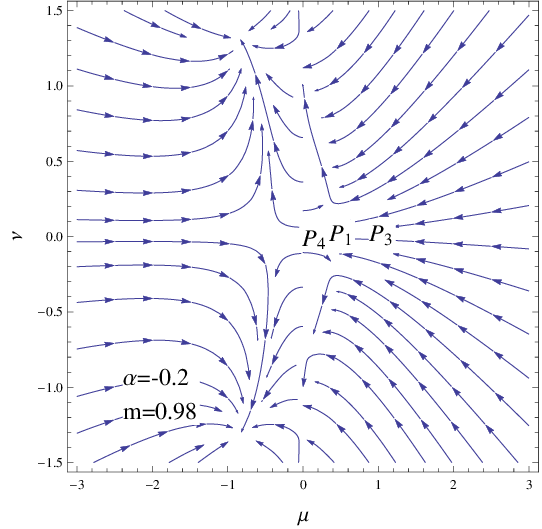,width=0.45\linewidth}\\
\epsfig{file=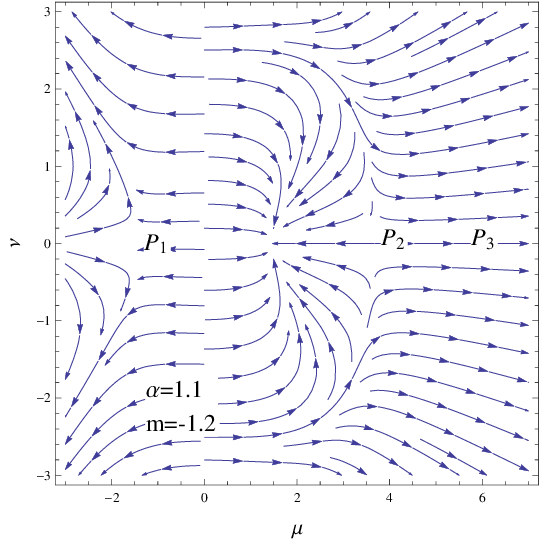,width=0.45\linewidth}\epsfig{file=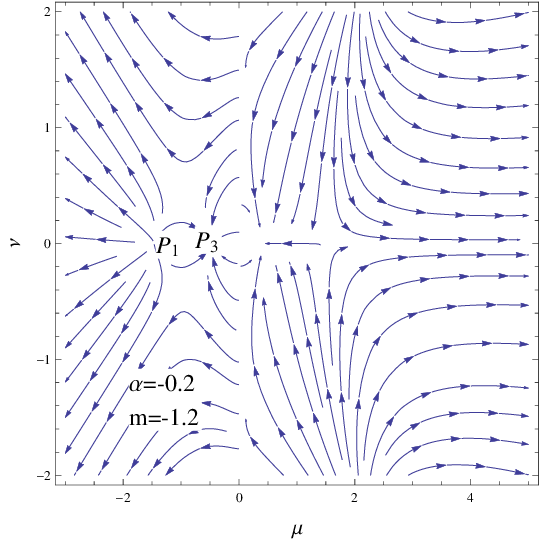,width=0.45\linewidth}
\caption{Plots for phantom coupled cosmic model with
$Q=\alpha\dot{\sigma}_m$ and $\lambda=2$.}
\end{figure}

For
$P_{2}=\left(\frac{-\sqrt{6}\lambda-\sqrt{6(\lambda^2+6)}}{2(2m+1)},
0\right)$, the corresponding eigenvalues are
\begin{eqnarray}\nonumber
\eta_{1}&=&\frac{9m}{m+2}\left[-1-\frac{3\alpha}{2(1-\alpha)(2m+1)}
\left\{\frac{-4(2m+1)^2}{[\sqrt{6}\lambda+\sqrt{6(\lambda^2+6)}]^2}
+\frac{2m+1}{3}\right\}\right.\\\label{11}&+&\left.\frac{3}{2(m+2)}
-\frac{[\sqrt{6}\lambda+\sqrt{6(\lambda^2+6)}]^2}{8(m+2)}\right],
\\\nonumber
\eta_2&=&-\frac{3m}{m+2}\left[\sqrt{\frac{3}{2}}\lambda
\frac{\sqrt{6(\lambda^2+6)\lambda}+\sqrt{6}\lambda}{m+2}
+\frac{1}{2(m+2)}\left\{-9\right.\right.\\\label{11a}&+&\left.\left.
\frac{[\sqrt{6(\lambda^2+6)}+\sqrt{6}\lambda]^2}{4}\right\}\right].
\end{eqnarray}
This point shows a stable attractor with decelerated expansion for
$m>0$. This point becomes unstable past attractor for negative
values of $m$. We find that $\alpha<0$ gives accelerated expansion
while the universe undergoes decelerated expansion for $\alpha=1.1$
as $q>0$.

For $P_{3}=\left(\frac{\sqrt{3(m+2)(2-\alpha)
+\sqrt{[3(m+2)(2-\alpha)]^2+36\alpha(1-\alpha)(m+2)}}}
{2(1-\alpha)(2m+1)}, 0\right)$, we have
\begin{eqnarray}\nonumber
\eta_{1}&=&\frac{9m}{m+2}\left[\frac{\alpha}{(m+2)(2-\alpha)
+\sqrt{[(m+2)(2-\alpha)]^2+4\alpha(1-\alpha)(m+2)}}
\right.\\\nonumber&+&\left.\frac{\alpha-2}{2(1-\alpha)}
+\frac{3}{2(m+2)}+\frac{2m+1}{2(1-\alpha)}\right.\\\nonumber&\times&\left.
\{3(m+2)(2-\alpha)+\sqrt{[3(m+2)(2-\alpha)]^2+36\alpha(1-\alpha)(m+2)}\}\right],
\\\label{12}\\\nonumber\eta_{2}&=&-\frac{3m}{m+2}\left[-\frac{9}{2(m+2)}
+\sqrt{\frac{3}{2}}\lambda\left(\frac{2m+1}{m+2}\right)
\right.\\\nonumber&\times&\left.\frac{\sqrt{3(m+2)(2-\alpha)
+\sqrt{[3(m+2)(2-\alpha)]^2+36\alpha(1-\alpha)(m+2)}}}
{2(1-\alpha)(2m+1)}\right.\\\nonumber&+&\left.
\frac{(2m+1)^2}{4(m+2)(2m+1)(1-\alpha)}\{3(m+2)(2-\alpha)
\right.\\\label{12a}&+&\left.\sqrt{[3(m+2)(2-\alpha)]^2
+36\alpha(1-\alpha)(m+2)}\}\right].
\end{eqnarray}
For positive values of $m$, this behaves as a saddle point showing
decelerated expansion ultimately followed by accelerated expansion.
For $m<0$, this point is saddle or stable showing accelerated
expanding universe model corresponding to $\alpha<0$ or $\alpha>0$,
respectively.\\ For $P_{4}=\left(\frac{\sqrt{3(m+2)(2-\alpha)
-\sqrt{[3(m+2)(2-\alpha)]^2+36\alpha(1-\alpha)(m+2)}}}
{2(1-\alpha)(2m+1)}, 0\right)$, we have
\begin{eqnarray}\nonumber
\eta_{1}&=&\frac{9m}{m+2}\left[\frac{\alpha}{(m+2)(2-\alpha)
-\sqrt{[(m+2)(2-\alpha)]^2+4\alpha(1-\alpha)(m+2)}}
\right.\\\nonumber&+&\left.\frac{\alpha-2}{2(1-\alpha)}+\frac{3}{2(m+2)}
+\frac{2m+1}{2(1-\alpha)}\right.\\\nonumber&\times&\left.\{3(m+2)(2-\alpha)
-\sqrt{[3(m+2)(2-\alpha)]^2+36\alpha(1-\alpha)(m+2)}\}\right],\\\label{13}
\\\nonumber\eta_{2}&=&-\frac{3m}{m+2}\left[-\frac{9}{2(m+2)}
+\sqrt{\frac{3}{2}}\lambda\left(\frac{2m+1}{m+2}\right)
\right.\\\nonumber&\times&\left.\frac{\sqrt{3(m+2)(2-\alpha)
-\sqrt{[3(m+2)(2-\alpha)]^2+36\alpha(1-\alpha)(m+2)}}}
{2(1-\alpha)(2m+1)}\right.\\\nonumber&+&\left.
\frac{(2m+1)^2}{4(m+2)(2m+1)(1-\alpha)}\{3(m+2)(2-\alpha)
\right.\\\label{13a}&-&\left.\sqrt{[3(m+2)(2-\alpha)]^2
+36\alpha(1-\alpha)(m+2)}\}\right].
\end{eqnarray}
This point behaves as a saddle point for all choices except for
$\alpha>0$ and $m>0$ that show stable future attractor.
\begin{table}[bht]
\textbf{Table 1:} \textbf{Phase Space Analysis for the Phantom
Coupled System with $Q=\alpha\dot{\sigma}_m$}. \vspace{0.5cm}
\centering
\begin{small}
\begin{tabular}{|c|c|c|c|c|c|c|c|}
\hline \textbf{Ranges of $m$ and $\alpha$ for Critical
Points}&\textbf{Stability}&\textbf{$\omega_{\phi}$}&\textbf{$q$}&\textbf{Acceleration}\\
\hline{\textbf{$P_1$}}&&&&\\
\hline{$m=0.98,~\alpha=1.1~(\alpha\neq1)$}&{Stable}&{-6.41}&{-1.87}&{Yes}\\
\hline{$m=0.98,~\alpha=-0.2$}&{Stable}&{-3.38}&{-0.73}&{Yes}\\
\hline{$m=-1.2,~\alpha=-0.2~(m\neq-2)$}&{Unstable}&{-1.18}&{-0.003}&{Yes}\\
\hline{$~m=-1.2,~\alpha=1.1$}&{Saddle}&{-3.92}&{-4.27}&{Yes}\\
\hline{\textbf{$P_2$}}&&&&\\
\hline{$m=0.98~,\alpha=1.1~(\alpha\neq1)$}&{Stable}&{1.301}&{1.096}&{No}\\
\hline{$m=0.98,~\alpha=-0.2$}&{Stable}&{-1.67}&{-1.23}&{Yes}\\
\hline{$m=-1.2,~\alpha=-0.2~(m\neq-0.5,-2)$}&{Unstable}&{-1.69}&{ -1.84}&{Yes}\\
\hline{$m=-1.2,~\alpha=1.1$}&{Unstable}&{1.3}&{6.8}&{No}\\
\hline{\textbf{$P_3$}}&&&&\\
\hline{$m=0.98,~\alpha=1.1(\alpha\neq1)$}&{Saddle}&{2.85}&{61.76}&{No}\\
\hline{$m=0.98,~\alpha=-0.2$}&{Saddle}&{-2.306}&{-1.383}&{Yes}\\
\hline{$m=-1.2,~\alpha=-0.2~(m\neq-0.5,-2)$}&{Saddle}&{-11.57}&{-5.56}&{Yes}\\
\hline{$m=-1.2,~\alpha=1.1$}&{Stable}&{2.16}&{46.25}&{No}\\
\hline{\textbf{$P_4$}}&&&&\\
\hline{$m=0.98,~\alpha=1.1~(\alpha\neq1)$}&{Stable}&{1.11}&{0.72}&{No}\\
\hline{$m=0.98,~\alpha=-0.2$}&{Saddle}&{-301.77}&{-2.49}&{Yes}\\
\hline{$m=-1.2,~\alpha=-0.2~(m\neq-0.5,-2)$}&{Saddle}&{-194.81}&{-6.54}&{Yes}\\
\hline{$m=-1.2,~\alpha=1.1$}&{Saddle}&{1.42}&{8.003}&{No}\\
\hline
\end{tabular}
\end{small}
\end{table}
It is mentioned here that all the points lie in accelerated
expanding phase of the universe for negative values of $\alpha$ as
$q<0$ and $\omega_{\phi}<-1$. We summarize the results for stability
of LRS BI universe coupled with phantom energy and matter in Table
\textbf{1}.

\subsection{Coupling $Q=\beta\dot{\sigma}_{\phi}$}

For this coupling, Eq.(\ref{7a}) leads to
\begin{eqnarray}\label{13}
\frac{\ddot{\phi}}{H\dot{\phi}}=-3-\sqrt{\frac{3}{2}}
\frac{2m+1}{m+2}\frac{\lambda\nu^2}{\mu}+\frac{3\beta}{1+\beta}.
\end{eqnarray}
Similarly, the autonomous system of equations takes the form
\begin{eqnarray}\nonumber
\mu'&=&\frac{3m}{m+2}\mu\left[-3-\sqrt{\frac{3}{2}}
\frac{2m+1}{m+2}\frac{\lambda\nu^2}{\mu}+\frac{3\beta
}{1+\beta}-\frac{1}{2(m+2)}\right.\\\label{13a}
&\times&\left.\{-9+(2m+1)^2(\mu^2+\nu^2)\}\right],\\\nonumber
\nu'&=&-\frac{3m}{m+2}\nu\left[\sqrt{\frac{3}{2}}
\frac{2m+1}{m+2}\mu\lambda+\frac{1}{2(m+2)}\{-9+(2m+1)^2
(\mu^2+\nu^2)\}\right],\\\label{13b}
\end{eqnarray}
We follow the same procedure to find the critical points. For
$P_1=(0,0)$, we have
\begin{eqnarray}\label{14}
\eta_1=-\frac{9m[2m+3\beta+7]}{2(m+2)^2(1+\beta)}, \quad
\eta_2=\frac{27m}{2(m+2)^2}.
\end{eqnarray}
This point gives a changing behavior of critical points
corresponding to different values of $\beta$ and $m$. For
$\beta=-1.2, 0.8$, the point $P_1$ is an unstable node by taking
$\lambda=2$ (Figure \textbf{2}). For $\beta=-1$, the eigenvalues
become undetermined, hence we neglect it. The negative value of $m$
gives stable future attractor. It is found that point $P_1$
indicates accelerated expansion of cosmos since $q<0$ for different
choices of parameters.
\begin{figure}\center
\epsfig{file=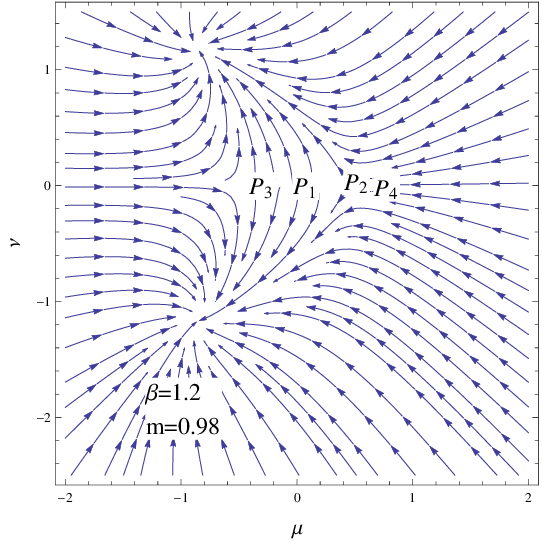,width=0.45\linewidth}\epsfig{file=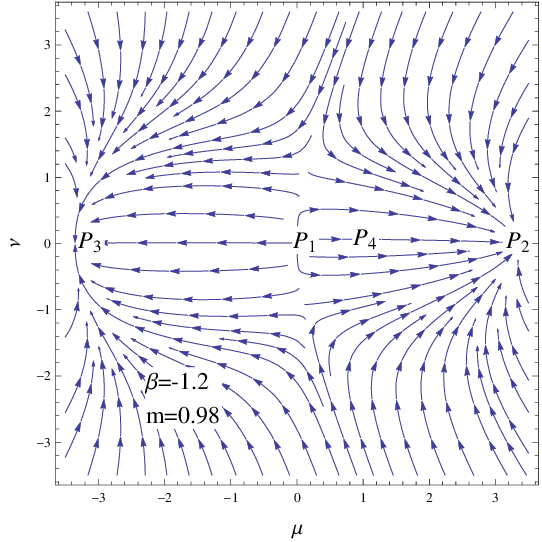,width=0.45\linewidth}\\
\epsfig{file=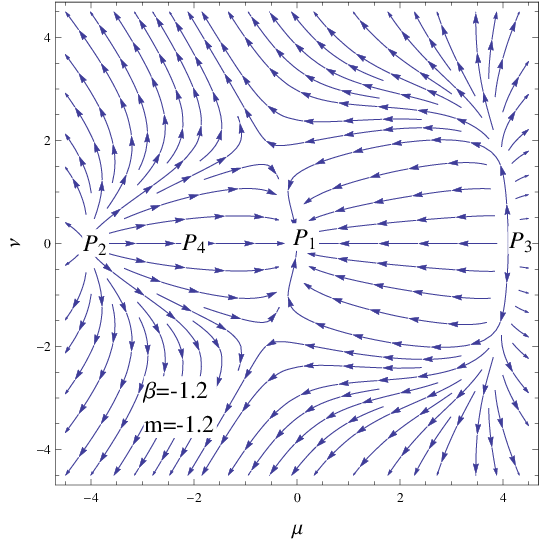,width=0.45\linewidth}\epsfig{file=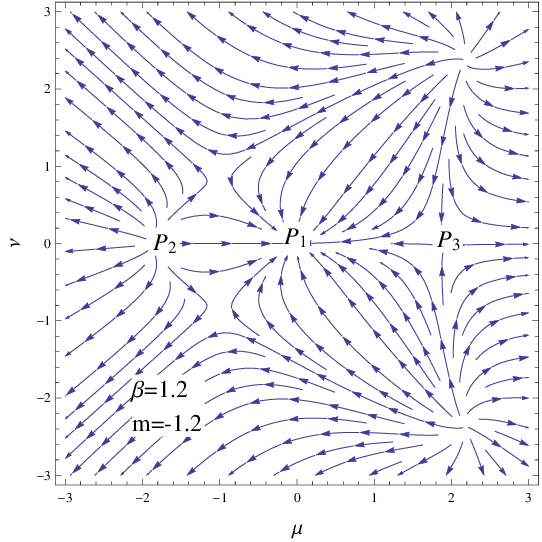,width=0.45\linewidth}
\caption{Plots for phantom coupled cosmic model with
$Q=\beta\dot{\sigma}_{\phi}$ and $\lambda=1.5$.}
\end{figure}

For $P_2=\left(\frac{1}{2m+1}\sqrt{\frac{3[3\beta-(2m+1)]}
{1+\beta}},0\right)$, the eigenvalues are given by
\begin{eqnarray}\label{14a}
\eta_1&=&\frac{9m}{(m+2)^2(1+\beta)}[3\beta-4m-5],
\\\label{14b}\eta_2&=&-\frac{9m}{(m+2)^2}\left[\lambda
\sqrt{\frac{3\beta-(2m+1)}{2(1+\beta)}}-\frac{m+2}{1+\beta}\right].
\end{eqnarray}
Here all values of $m$ and $\beta$ correspond to saddle or unstable
points except $m>0$ and $\beta<0$ that yield stable future
attractor. The universe undergoes decelerated expansion for negative
values of $\beta$ while $\beta>0$ corresponds to the accelerated
expansion. For
$P_3=\left(-\frac{1}{2m+1}\sqrt{\frac{3[3\beta-(2m+1)]}{1+\beta}},0\right)$,
the eigenvalues are obtained as
\begin{eqnarray}\label{14c}
\eta_1&=&\frac{9m}{(m+2)^2(1+\beta)}[3\beta-4m-5],
\\\label{14d}\eta_2&=&\frac{9m}{(m+2)^2}\left[\lambda
\sqrt{\frac{3\beta-(2m+1)}{2(1+\beta)}}+\frac{m+2}{1+\beta}\right].
\end{eqnarray}
We find that point $P_{3}$ is a saddle node for all choices of
$\beta$ and $m$ except for $\beta<0$ and $m>0$ at which it behaves
as a stable point. This point yields decelerated expansion for
negative values of $\beta$. For
$P_4=\left(0,\pm\frac{3}{2m+1}\right)$, the eigenvalues are
\begin{eqnarray}\label{15a}
\eta_1=-\frac{9m}{(m+2)(1+\beta)},\quad \eta_2=-\frac{27m}{(m+2)^2}.
\end{eqnarray}
This point is a saddle node for negative values of $\beta$ while
$\beta>0$ corresponds to stable/unstable attractors depending on
parameter $m$. It is found that $P_4$ lies in a region of
accelerated expansion for all the choices of parameters. Table
\textbf{2} shows the dynamical analysis for LRS BI model.
\begin{table}[bht]
\textbf{Table 2:} \textbf{Phase Space Analysis for the Phantom
Coupled System with $Q=\beta\dot{\sigma}_{\phi}$}. \vspace{0.5cm}
\centering
\begin{small}
\begin{tabular}{|c|c|c|c|c|c|c|c|}
\hline \textbf{Ranges of $m$ and $\beta$ for Critical
Points}&\textbf{Stability}&\textbf{$\omega_{\phi}$}&\textbf{$q$}&\textbf{Acceleration}\\
\hline{\textbf{$P_1$}}&&&&\\
\hline{$m=0.98,~\beta=1.2$}&{Saddle}&{Undetermined}&{-2.51}&{Yes}\\
\hline{$m=0.98,~\beta=-1.2,~(\beta\neq-1)$}&{Unstable}&{Undetermined}&{-2.51}&{Yes}\\
\hline{$m=-1.2,~\beta=-1.2$}&{Stable}&{Undetermined}&{-6.63}&{Yes}\\
\hline{$m=-1.2,~\beta=1.2$}&{Stable}&{Undetermined}&{-6.63}&{Yes}\\
\hline{\textbf{$P_2$}}&&&&\\
\hline{$m=0.98,~\beta=1.2$}&{Saddle}&{-37.41}&{-2.36}&{Yes}\\
\hline{$m=0.98,~\beta=-1.2,~(\beta\neq-1)$}&{Stable}&{2.58}&{14}&{No}\\
\hline{$m=-1.2,~\beta=-1.2,~(m\neq-0.5)$}&{Saddle}&{1.72}&{14}&{No}\\
\hline{$m=-1.2,~\beta=1.2$}&{Unstable}&{-1.07}&{-2.36}&{Yes}\\
\hline{\textbf{$P_3$}}&&&&\\
\hline{$m=0.98,~\beta=1.2$}&{Saddle}&{-37.42}&{-2.36}&{Yes}\\
\hline{$m=0.98,~\beta=-1.2,~(\beta\neq-1)$}&{Stable}&{2.58}&{14}&{No}\\
\hline{$m=-1.2,~\beta=-1.2,~(m\neq-0.5)$}&{Saddle}&{1.72}&{14}&{No}\\
\hline{$m=-1.2,~\beta=1.2$}&{Saddle}&{-1.07}&{-2.36}&{Yes}\\
\hline{\textbf{$P_4$}}&&&&\\
\hline{$m=0.98,~\beta=1.2$}&{Stable}&{-0.97}&{-1}&{Yes}\\
\hline{$m=0.98,~\beta=-1.2,~(\beta\neq-1)$}&{Saddle}&{-0.97}&{-1}&{Yes}\\
\hline{$m=-1.2,~\beta=-1.2$}&{Saddle}&{-0.217}&{-1}&{Yes}\\
\hline{$m=-1.2,~\beta=1.2$}&{Unstable}&{-0.217}&{-1}&{Yes}\\
\hline
\end{tabular}
\end{small}
\end{table}

\subsection{Coupling
$Q=\gamma(\dot{\sigma}_{m}+\dot{\sigma}_{\phi})$}

In this section, we assume a combination of the form
$\dot{\sigma}_{m}$ and $\dot{\sigma}_{\phi}$ for the coupling
through which Eq.(\ref{7a}) becomes
\begin{eqnarray}\label{13}
\frac{\ddot{\phi}}{H\dot{\phi}}=-3-\sqrt{\frac{3}{2}}
\frac{2m+1}{m+2}\frac{\lambda\nu^2}{\mu}
-\frac{3\gamma\Omega_m}{2(1-\gamma)\mu^2}+\frac{3\gamma}{1+\gamma}.
\end{eqnarray}
The evolution and conservation equations are given by
\begin{eqnarray}\nonumber
\mu'&=&\frac{3m}{m+2}\mu\left[-3-\sqrt{\frac{3}{2}}
\frac{2m+1}{m+2}\frac{\lambda\nu^2}{\mu}-\frac{3\gamma
\Omega_m}{2(1-\gamma)\mu^2}+\frac{3\gamma}{1+\gamma}
-\frac{1}{2(m+2)}\right.\\\label{9a}&\times&\left.
\{-9+(2m+1)^2(\mu^2+\nu^2)\}\right],\\\nonumber
\nu'&=&-\frac{3m}{m+2}\nu\left[\sqrt{\frac{3}{2}}\mu\lambda
\left(\frac{2m+1}{m+2}\right)+\frac{1}{2(m+2)}
\{-9+(2m+1)^2(\mu^2+\nu^2)\}\right].\\\label{9b}
\end{eqnarray}
For $P_{1}=\left(\frac{-\frac{2m+1}{m+2}\sqrt{\frac{3}{2}}\lambda
+\sqrt{\frac{3}{2}\lambda^2+\left(\frac{3(2m+1)}{m+2}\right)^2}}
{(2m+1)^2/(m+2)},0\right)$,
the corresponding eigenvalues are
\begin{eqnarray}\nonumber
\eta_1&=&\frac{9m}{m+2}\left[\frac{1-2m}{2(m+2)}+\frac{\gamma}{2(1-\gamma)}
\left\{\frac{(2m+1)^4}{(m+2)^2\left\{-\frac{2m+1}{m+2}\sqrt{\frac{3}{2}}\lambda
+\sqrt{\frac{3}{2}\lambda^2+\left(\frac{3(2m+1)}{m+2}\right)^2}\right\}^2}
\right.\right.\\\label{9c}&-&\left.\left.\frac{2m+1}{3}\right\}
+\frac{\gamma}{1+\gamma}-\frac{1}{2(m+2)}\left\{\frac{-\frac{2m+1}{m+2}
\sqrt{\frac{3}{2}}\lambda+\sqrt{\frac{3}{2}\lambda^2+\left(\frac{3(2m+1)}
{m+2}\right)^2}}{(2m+1)^2/(m+2)}\right\}^2\right],\\\nonumber
\eta_2&=&-\frac{3m}{m+2}\left[\sqrt{\frac{3}{2}}\lambda
\left\{\frac{-\frac{2m+1}{m+2}\sqrt{\frac{3}{2}}\lambda+\sqrt{\frac{3}{2}\lambda^2
+\left(\frac{3(2m+1)}{m+2}\right)^2}}{(2m+1)}\right\}+\frac{1}{2(m+2)}
\right.\\\nonumber&\times&\left.\left\{-9+\left(\frac{m+2}{2m+1}\right)^2
\left[-\left(\frac{2m+1}{m+2}\right)\sqrt{\frac{3}{2}}\lambda
+\sqrt{\frac{3}{2}\lambda^2+\left(\frac{3(2m+1)}{m+2}\right)^2}\right]^2\right\}
\right].\\\label{b1}
\end{eqnarray}
In this case, we examine unstable nodes for $m<0$ by varying
$\gamma$ that shows decelerated expanding universe except for
$m=-0.5, -2$ (Figure \textbf{3}). On the other hand, the positive
values of $m$ give accelerated expansion of the universe.
\begin{figure}\center
\epsfig{file=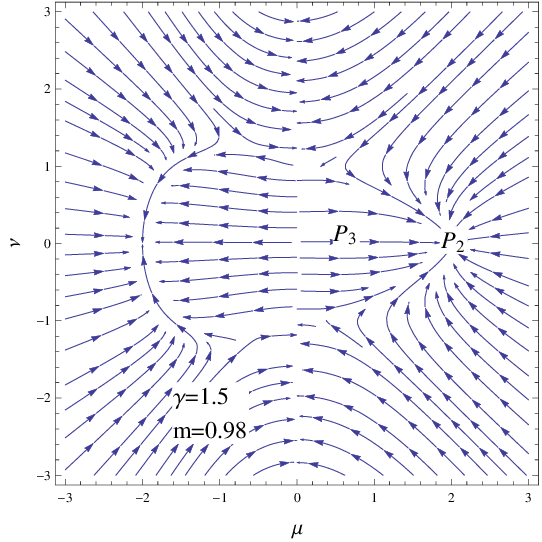,width=0.45\linewidth}\epsfig{file=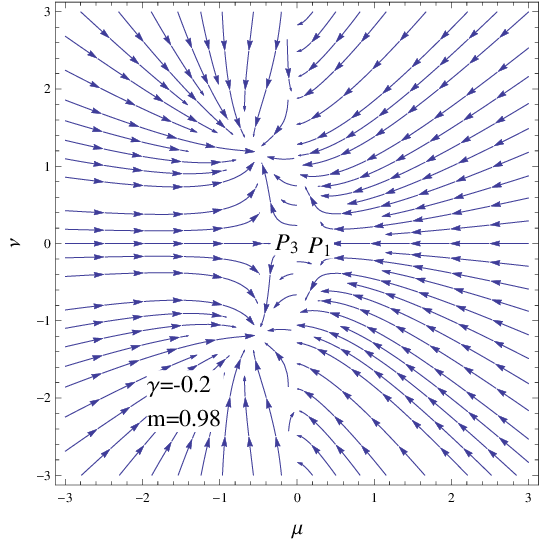,width=0.45\linewidth}\\
\epsfig{file=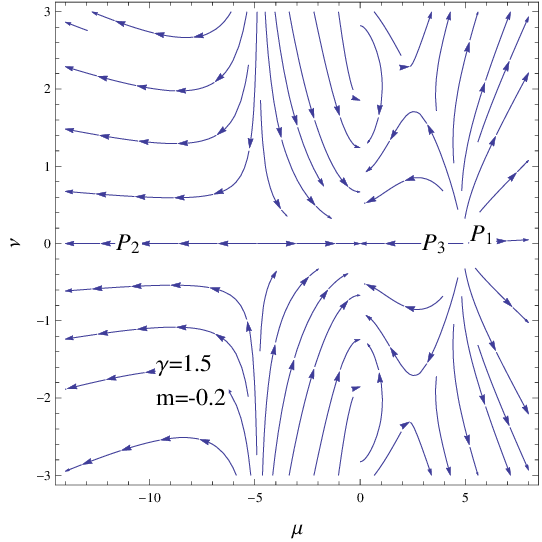,width=0.45\linewidth}\epsfig{file=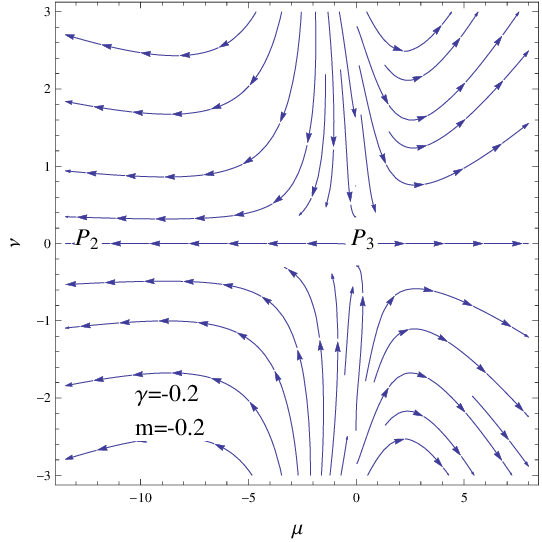,width=0.45\linewidth}
\caption{Plots for phantom coupled cosmic model with
$Q=\gamma(\dot{\sigma}_{m}+\dot{\sigma}_{\phi})$ and $\lambda=1.5$.}
\end{figure}

For $P_{2}=\left(\frac{-\frac{2m+1}{m+2}\sqrt{\frac{3}{2}}\lambda
-\sqrt{\frac{3}{2}\lambda^2+\left(\frac{3(2m+1)}{m+2}\right)^2}}{(2m+1)^2/(m+2)},0\right)$,
the corresponding eigenvalues are given as
\begin{eqnarray}\nonumber
\eta_1&=&\frac{9m}{m+2}\left[\frac{1-2m}{2(m+2)}+\frac{\gamma}{2(1-\gamma)}
\left\{\frac{(2m+1)^4}{(m+2)^2\left\{-\frac{2m+1}{m+2}\sqrt{\frac{3}{2}}\lambda
-\sqrt{\frac{3}{2}\lambda^2+\left(\frac{3(2m+1)}{m+2}\right)^2}\right\}^2}
\right.\right.\\\label{a2}&-&\left.\left.\frac{2m+1}{3}\right\}
+\frac{\gamma}{1+\gamma}-\frac{1}{2(m+2)}\left\{\frac{-\frac{2m+1}{m+2}
\sqrt{\frac{3}{2}}\lambda-\sqrt{\frac{3}{2}\lambda^2
+\left(\frac{3(2m+1)}{m+2}\right)^2}}{(2m+1)^2/(m+2)}\right\}^2\right],
\\\nonumber\eta_2&=&-\frac{3m}{m+2}\left[\sqrt{\frac{3}{2}}\lambda
\left\{\frac{-\frac{2m+1}{m+2}\sqrt{\frac{3}{2}}\lambda-\sqrt{\frac{3}{2}\lambda^2
+\left(\frac{3(2m+1)}{m+2}\right)^2}}{(2m+1)}\right\}
\right.\\\nonumber&+&\left.\frac{1}{m+2}\left\{-9+\frac{(m+2)^2}{m+2}
\left[-\left(\frac{2m+1}{m+2}\right)\frac{3}{2}\lambda-\sqrt{\frac{3}{2}\lambda^2
+\left(\frac{3(2m+1)}{m+2}\right)^2}\right]^2\right\}
\right].\\\label{b2}
\end{eqnarray}
This point corresponds to either stable or unstable nodes for $m>0$
or $m<0$, respectively. It is found that $P_{2}$ lies in the matter
dominated phase for all choices of the parameters. For
\\$P_{3}=\left(\frac{\sqrt{m+2}}{(2m+1)}\sqrt{\frac{\gamma(2m+1)}{2(1-\gamma)}-\frac{3}{1+\gamma}
+\frac{9}{2(m+2)}+\xi},0\right)$, the eigenvalues are
\begin{eqnarray}\nonumber
\eta_1&=&\frac{9m}{m+2}\left[\frac{2(2m+1)^2\gamma(1+\gamma)}{2\gamma(m+2)
+4(m+1)(1-\gamma^2)(\xi-3)+9}+\frac{1}{2(1+\gamma)}\right.
\\\label{c1}&+&\left.\frac{5\gamma(2m+1)}{12(1-\gamma)}-\frac{3}{4(m+2)}
-\frac{\xi}{2}\right],\\\label{d1}\eta_2&=&-\frac{3m}{m+2}
\left[\sqrt{\frac{3}{2(m+2)}\left(\frac{\gamma(2m+1)}{2(1-\gamma)}
-\frac{3}{1+\gamma}+\frac{9}{2(m+2)}+\xi\right)}\right.\\\nonumber
&+&\left.\frac{1}{2(m+2)}\left\{-9+(m+2)\left(\frac{\gamma(2m+1)}{2(1-\gamma)}
-\frac{3}{1+\gamma}+\frac{9}{2(m+2)}+\xi\right)\right\}\right],\\\label{d2}
\end{eqnarray}
where
\begin{eqnarray}\nonumber
\xi&=&\sqrt{\frac{3}{1+\gamma}-\frac{\gamma(2m+1)}{2(1-\gamma)}
-\frac{9}{2(m+2)}-\frac{3(2m+1)^2\gamma}{(m+2)(1-\gamma)}}.
\end{eqnarray}
\begin{table}[bht]
\textbf{Table 3:} \textbf{Phase Space Analysis for
$Q=\gamma(\dot{\sigma}_{m}+\dot{\sigma}_{\phi})$}. \vspace{0.5cm}
\centering
\begin{small}
\begin{tabular}{|c|c|c|c|c|c|c|c|}
\hline \textbf{Ranges of $m$ and $\gamma$ for Critical
Points}&\textbf{Stability}&\textbf{$\omega_{\phi}$}&\textbf{$q$}&\textbf{Acceleration}\\
\hline{\textbf{$P_1$}}&&&&\\
\hline{$m=0.98,~\gamma=1.5,~(\gamma\neq1)$}&{Stable}&{-14.18}&{-2.16}&{Yes}\\
\hline{$m=0.98,~\gamma=-0.2,~(\gamma\neq-1)$}&{Saddle}&{-14.18}&{-2.16}&{Yes}\\
\hline{$m=-0.2,~\gamma=-0.2,~(m\neq-0.5,-2)$}&{Unstable}&{0.128}&{4.86}&{No}\\
\hline{$m=-0.2,~\gamma=1.5$}&{Unstable}&{0.128}&{4.86}&{No}\\
\hline{\textbf{$P_2$}}&&&&\\
\hline{$m=0.98,~\gamma=1.5,~(\gamma\neq1)$}&{Stable}&{2.06}&{4.22}&{No}\\
\hline{$m=0.98,~\gamma=-0.2,~(\gamma\neq-1)$}&{Stable}&{2.06}&{4.22}&{No}\\
\hline{$m=-0.2,~\gamma=-0.2,~(m\neq-0.5,-2)$}&{Unstable}&{0.179}&{26.46}&{No}\\
\hline{$m=-0.2,~\gamma=1.5$}&{Unstable}&{0.179}&{26.46}&{No}\\
\hline{\textbf{$P_3$}}&&&&\\
\hline{$m=0.98,~\gamma=1.5,~(\gamma\neq1)$}&{Saddle}&{57.94}&{-2.62}&{Yes}\\
\hline{$m=0.98,~\gamma=-0.2,~(\gamma\neq-1)$}&{Saddle}&{57.94}&{-2.62}&{Yes}\\
\hline{$m=-0.2,~\gamma=-0.2,~(m\neq-2)$}&{Saddle}&{-1.504}&{-3.503}&{Yes}\\
\hline{$m=-0.2,~\gamma=1.5$}&{Saddle}&{-1.504}&{-3.503}&{Yes}\\
\hline
\end{tabular}
\end{small}
\end{table}
The behavior of eigenvalues as well as trajectories indicate that
point $P_3$ is a saddle node for all choices of the parameters. The
summary of the above results is provided in Table \textbf{3}.

\section{Coupled Tachyon Dynamics}

Now we study the evolution of LRS BI universe by considering tachyon
coupled cosmic fluid. The conservation equations are
\begin{eqnarray}\label{30}
\dot{\sigma}_m+3(\sigma_m+p_m)H&=&Q,\\\label{30a}
\dot{\sigma}_{\phi}+3(\sigma_{\phi}+p_{\phi})H&=&-Q,
\end{eqnarray}
where $\sigma_{\phi}=\frac{V(\phi)}{\sqrt{1-\dot{\phi}^2}}$ and
$p_{\phi}=-V(\phi)\sqrt{1-\dot{\phi}^2}$. In this case, the
evolution equations take the form
\begin{eqnarray}\label{31}
H^2&=&\frac{(m+2)^2}{9(2m+1)}\left[\frac{V(\phi)}
{\sqrt{1-\dot{\phi}^2}}+\sigma_m\right], \\\label{31a}
\frac{\ddot{\phi}}{1-\dot{\phi}^2}&=&
-\left[3H\dot{\phi}+\frac{V'(\phi)}{V(\phi)}
+\frac{Q\sqrt{1-\dot{\phi}^2}}{V(\phi)\dot{\phi}}\right].
\end{eqnarray}
Here we define the following dimensionless quantities
\begin{eqnarray}\label{32}
\mu=\frac{(m+2)\dot{\phi}}{2m+1}, \quad
\nu=\frac{(m+2)\sqrt{V}}{\sqrt{3}(2m+1)H}, \quad
\lambda=-\frac{V'}{V\sqrt{V}},
\end{eqnarray}
such that the dynamical system yields
\begin{eqnarray}\label{33}
\mu'&=&\frac{3m}{2m+1}\frac{\ddot{\phi}\mu}{H\dot{\phi}},
\\\label{33a}\nu'&=&-\frac{3\sqrt{3}m(2m+1)^2\lambda\mu\nu^2}{2(m+2)^3}
-\frac{3m\nu}{m+2}\frac{\dot{H}}{H^2}.
\end{eqnarray}
We consider inverse square potential with constant parameter
$\lambda$. In this case, we take the only coupling
$Q=\beta\dot{\sigma}_{\phi}$ through which Eqs.(\ref{31}) and
(\ref{31a}) lead to
\begin{eqnarray}\label{34}
\frac{\dot{H}}{H^2}&=&\frac{(2m+1)^2}{2(m+2)}\nu^2\sqrt{1
-\left(\frac{2m+1}{m+2}\right)^2\mu^2}-\frac{9}{2(m+2)},
\\\label{34a}\frac{\ddot{\phi}}{\dot{\phi}H}&=&\left[1-\left
(\frac{2m+1}{m+2}\right)^2\mu^2\right]\left[\frac{\sqrt{3}
\nu\lambda}{\mu}+\frac{3\beta}{1+\beta}-3\right].
\end{eqnarray}
The effective EoS and deceleration parameters can be written as
\begin{eqnarray}\label{35}
w_{eff}&=&-1-\frac{1}{3(m+2)}\left[(2m+1)^2\nu^2
\sqrt{1-\left(\frac{2m+1}{m+2}\right)^2\mu^2}-9\right],
\\\label{35a}q&=&-1-\frac{1}{2(m+2)}\left[(2m+1)^2\nu^2
\sqrt{1-\left(\frac{2m+1}{m+2}\right)^2\mu^2}-9\right].
\end{eqnarray}
\begin{figure}\center
\epsfig{file=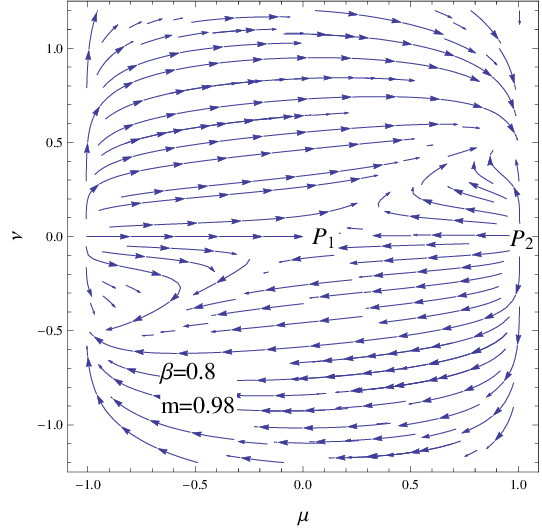,width=0.45\linewidth}\epsfig{file=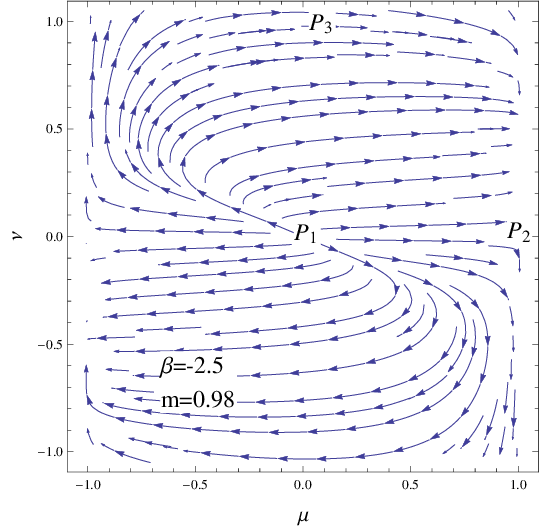,width=0.45\linewidth}\\
\epsfig{file=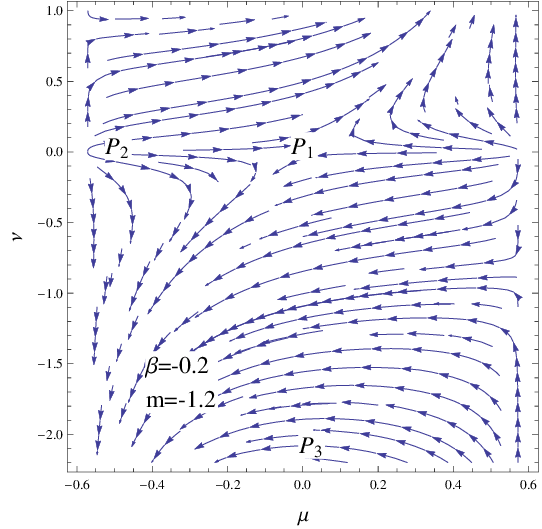,width=0.45\linewidth}\epsfig{file=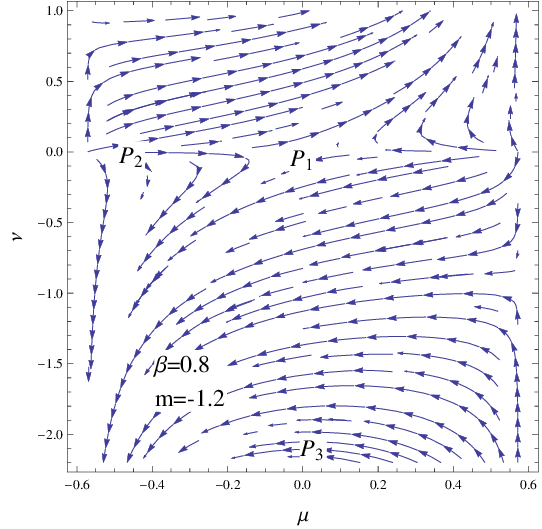,width=0.45\linewidth}
\caption{Plots for tachyon coupled cosmic model with
$Q=\beta\dot{\sigma}_{\phi}$ and $\lambda=2$.}
\end{figure}

For $P_1=(0, 0)$, we find the eigenvalues as
\begin{eqnarray}\label{36b}
\eta_1=-\frac{9m}{(2m+1)(1+\beta)}, \quad
\eta_2=\frac{27m}{2(m+2)^2}.
\end{eqnarray}
We find that point $P_1$ is saddle/unstable for $m>$ which becomes
stable for negative values of $m$ ($m\neq-0.5,\beta\neq-1$)
undergoing accelerated expansion as $q<0$ (Figure \textbf{4}). We
provide the obtained results in Table \textbf{4}.
\begin{table}[bht]
\textbf{Table 4:} \textbf{Phase Space Analysis of Tachyon Coupled
System with $Q=\beta\dot{\sigma}_{\phi}$}.
\vspace{0.5cm}
\centering
\begin{small}
\begin{tabular}{|c|c|c|c|c|c|c|c|}
\hline \textbf{Ranges of $m$ and $\beta$ for Critical
Points}&\textbf{Stability}&\textbf{$\omega_{\phi}$}&\textbf{$q$}&\textbf{Acceleration}\\
\hline{\textbf{$P_1$}}&&&&\\
\hline{$m=0.98,~\beta=0.8$}&{Saddle}&{-2.006}&{-2.51}&{Yes}\\
\hline{$m=0.98,~\beta=-0.2,(~\beta\neq-1)$}&{Unstable}&{-2.006}&{-2.51}&{Yes}\\
\hline{$m=-1.2,~\beta=-0.2,~(m\neq-0.5)$}&{Stable}&{-4.75}&{-6.625}&{Yes}\\
\hline{$m=-1.2,~\beta=0.8$}&{Stable}&{-4.75}&{-6.625}&{Yes}\\
\hline{\textbf{P$_2$}}&&&&\\
\hline{$m=0.98,~\beta=0.8$}&{Unstable}&{-2.006}&{-2.51}&{Yes}\\
\hline{$m=0.98,~\beta=-0.2,~(\beta\neq-1)$}&{Saddle}&{-2.006}&{-2.51}&{Yes}\\
\hline{$m=-1.2,~\beta=-0.2,~(m\neq-0.5)$}&{Saddle}&{-4.75}&{-6.625}&{Yes}\\
\hline{$m=-1.2,~\beta=0.8$}&{Saddle}&{-4.75}&{-6.625}&{Yes}\\
\hline{\textbf{P$_3$}}&&&&\\
\hline{$m=0.98,~\beta=0.8$}&{Stable}&{-1}&{-1}&{Yes}\\
\hline{$m=0.98,~\beta=-0.2,~(\beta\neq-1)$}&{Saddle}&{-1}&{-1}&{Yes}\\
\hline{$m=-1.2,~\beta=-0.2,~(m\neq-0.5)$}&{Saddle}&{-1}&{-1}&{Yes}\\
\hline{$m=-1.2,~\beta=0.8$}&{Saddle}&{-1}&{-1}&{Yes}\\
\hline
\end{tabular}
\end{small}
\end{table}

For $P_2=\left(\pm\frac{m+2}{2m+1}, 0\right)$, the eigenvalues
become
\begin{eqnarray}\label{36c}
\eta_1=\frac{18m}{(2m+1)(1+\beta)}, \quad
\eta_2=\frac{27m}{2(m+2)^2},
\end{eqnarray}
which correspond to unstable nodes for $m,\beta>0$ lying in
accelerated expanding phase of the universe model. We have saddle
nodes for the remaining choices of parameters $m$ and $\beta$. The
cosmic portrait indicates accelerated expansion phase of the
universe for all choices of the parameters. When $P_3=\left(0,
\pm\frac{3}{2m+1}\right)$, the eigenvalues yield
\begin{eqnarray}\label{36b}
\eta_1=-\frac{9m}{(2m+1)(1+\beta)}, \quad
\eta_2=-\frac{27m}{(m+2)^2}.
\end{eqnarray}
This point is a stable future attractor for $\beta>0$ and $m>0$
undergoing accelerated expansion of the cosmic model. We find that
the respective point corresponds to saddle nodes for all other
choices of parameters except at $m=-0.5, \beta-1=$. This point
indicates de Sitter ($q=-1$, $w_{eff}=-1$) phase of cosmos.

\section{Power-Law Scale Factor}

Here we explore the power-law behavior of the scale factor by
applying some assumptions to both the phantom and tachyon coupled
fields. In this context, Eq.(\ref{7}) becomes
\begin{equation}\label{50}
\dot{\Theta}=\frac{1}{6(m+2)}[-9+(2m+1)^2(\mu^2+\nu^2)]\Theta^2,
\end{equation}
where $\Theta=3H$ is the expansion scalar. For $\Theta\neq0$, we
formulate power-law scale factor only if
$-9+(2m+1)^2(\mu^2+\nu^2)\neq0$. We obtain the respective generic
critical point by solving
$\Theta=\frac{\dot{a}}{a}+\frac{2\dot{b}}{b}$ for $a(t)$ and $b(t)$
as
\begin{equation}\label{51}
b^{(m+2)}=b_{0}^{(m+2)}(t-t_{0})^{\frac{-6(m+2)}{-9+(2m+1)^2(\mu^2+\nu^2)}}.
\end{equation}
It is noticed that behavior of the term
$``-9+(2m+1)^2(\mu^2+\nu^2)"$ is quite important to assess different
cosmological phases. It shows exponential cosmic expansion for
$-9+(2m+1)^2(\mu^2+\nu^2)=0$ while
$-9+(2m+1)^2(\mu^2+\nu^2)\gtrless0$ gives accelerated cosmic
expansion or contraction, respectively. Figure \textbf{5} depicts
different states for power-law scale factor, where red and gray
regions show contraction and accelerated expansion of cosmic model,
respectively. It is found that decelerated expansion tends to
increase by increasing $m$ whereas $m<0$ undergoes decelerated
expansion.
\begin{figure}\center
\epsfig{file=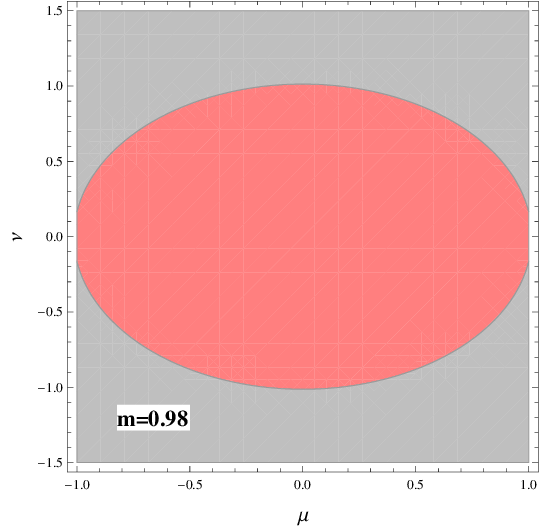,width=0.45\linewidth}\epsfig{file=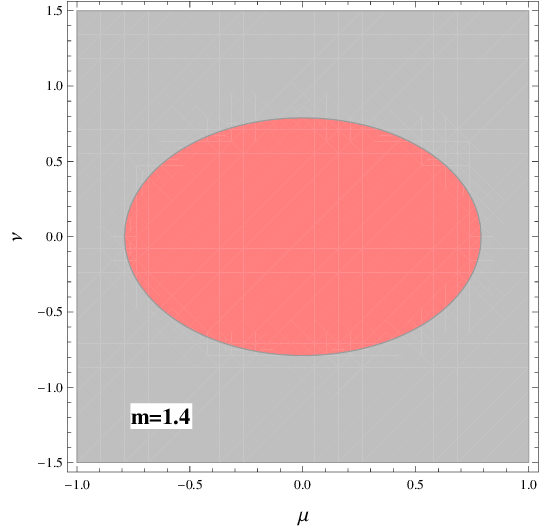,width=0.45\linewidth}
\epsfig{file=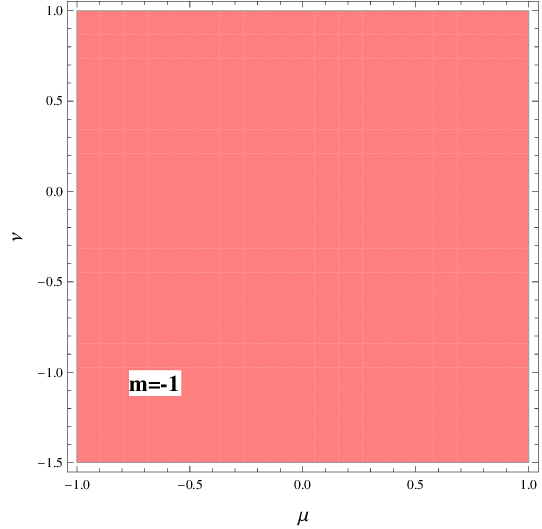,width=0.45\linewidth}\caption{Plots for power-law
scale factor coupled with phantom fluid. Blue and gray zones
represent contraction and accelerated expansion of cosmos,
respectively.}
\end{figure}

For tachyon coupled fluid, Eq.(\ref{34}) becomes
\begin{equation}\label{34aa}
\dot{\Theta}=\frac{1}{6(m+2)}\left[(2m+1)^2\nu^2\sqrt{1
-\left(\frac{2m+1}{m+2}\right)^2\mu^2}-\frac{9}{2(m+2)}\right]\Theta^2.
\end{equation}
For $\Theta\neq0$, we again find power-law scale factor if
$(2m+1)^2\nu^2\sqrt{1
-\left(\frac{2m+1}{m+2}\right)^2\mu^2}-\frac{9}{2(m+2)}\neq0$. The
generic critical point is obtained through
$\Theta=\frac{\dot{a}}{a}+\frac{2\dot{b}}{b}$ as
\begin{equation}\label{51a}
b^{(m+2)}=b_{0}^{(m+2)}(t-t_{0})^{\frac{-6(m+2)}{(2m+1)^2\nu^2\sqrt{1
-\left(\frac{2m+1}{m+2}\right)^2\mu^2}-\frac{9}{2(m+2)}}}.
\end{equation}
We examine various cosmological states according to
$(2m+1)^2\nu^2\sqrt{1
-\left(\frac{2m+1}{m+2}\right)^2\mu^2}-\frac{9}{2(m+2)}\gtrless0$.
Figure \textbf{6} corresponds to the different cosmological phases,
where blue and gray zones indicate contraction as well as
accelerated expansion of cosmos, respectively. It is found that the
region for decelerated expansion decreases by increasing $m$ whereas
$m<0$ corresponds to contraction of the cosmic model (Figure
\textbf{6}).
\begin{figure}\center
\epsfig{file=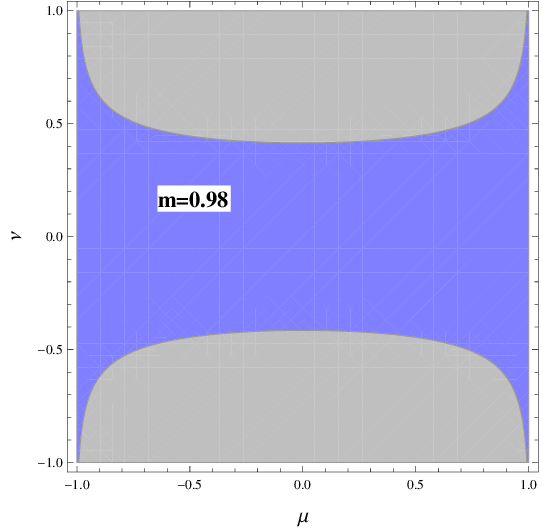,width=0.45\linewidth}\epsfig{file=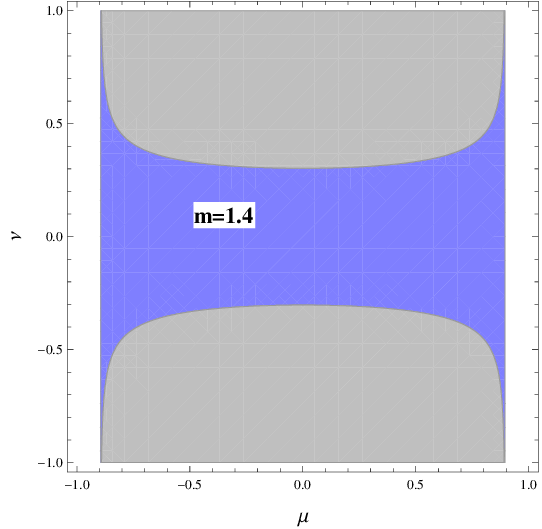,width=0.45\linewidth}
\epsfig{file=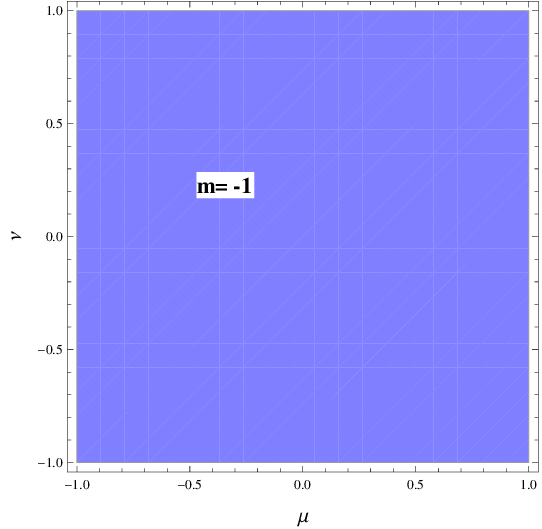,width=0.45\linewidth}\caption{Plots of
qualitative phase space analysis for power-law scale factor with
tachyon coupled matter.}
\end{figure}

\section{Summary}

This work is devoted to discuss stability of LRS BI universe model
by taking scalar field coupled fluid. An autonomous system of
equations has been developed by introducing various dimensionless
parameters. We have obtained the corresponding critical points with
different choices of the parameters. The eigenvalues have been
formulated which characterize the obtained critical points. We
summarize the results of the dynamical analysis as follows.

Firstly, we have studied stability of the universe model dominated
by the coupling of phantom energy and DM for different values of $m$
(Figures \textbf{1-3}). We have considered three different linear
combinations for the coupling parameter. For
$Q=\alpha\dot{\sigma}_m$ and $(P_1)$, we have observed stable DE
dominated state with $m>0$ and different choices of $\alpha$ (Figure
\textbf{1}). This analysis indicates a deceleration phase eventually
followed by accelerated expansion of the cosmos for point $P_3$. It
is observed that point $P_2$ lies in non-accelerating phase of the
universe as $q>0$ for $\alpha>0$ which may alleviate coincidence
problem as compared to \cite{24a,27}. For $m>0$, all the points show
stable future attractors except $P_3$ and $P_4$ which shows saddle
nodes depending on the choices of $\alpha$.

For $Q=\beta\dot{\sigma}_{\phi}$, all the eigenvalues and
trajectories show saddle nodes for positive values of $m,\beta$
(except $P_4$) which become stable/unstable for $m<0$ corresponding
to different choices of $\beta$ (Figure \textbf{2}). It is noted
that $P_1$ and $P_4$ undergo accelerated expansion for all choices
of the parameters. For the coupling
$Q=\gamma(\dot{\sigma}_{m}+\dot{\sigma}_{\phi})$, we have found
$P_1$ as stable attractor for positive values of $m$ showing
accelerated cosmic expansion. When $m<0$, unstable node is observed
for both $P_1$ and $P_2$ undergoing decelerated expansion (Figure
\textbf{3}). In this case, point $P_3$ shows saddle node in
accelerating phases. For the coupling $Q=\beta\dot{\sigma}_{\phi}$,
$P_3$ is the only point that indicates accelerated expanding
universe for all the chosen values of parameters.

Secondly, we have discussed dynamics of the universe model by
considering tachyon coupled field. In this case, we consider
$Q=\beta\dot{\sigma}_{\phi}$ only. The cosmic portrait indicates
accelerated expansion of the universe model (Figure \textbf{4}). For
$m>0$, points $P_1$ and $P_2$ behave as unstable/saddle nodes
whereas $P_3$ provides stable future attractor in accelerated
expansion scenario. For $m<0$, the point $P_1$ is stable showing
accelerated expansion of cosmos while the remaining points act as
saddle nodes that correspond to de Sitter cosmic phase. We conclude
that tachyon coupled field yields an era of accelerated expansion
for all choices of parameters.

Finally, we have investigated the characteristics of power-law scale
factor by varying $m$. It indicates various phases of evolution
(accelerated or exponential expansion) for the respective cosmic
model as shown (Figures \textbf{5} and \textbf{6}). In case of
phantom coupled matter, the region of decelerated expansion
increases by increasing $m$ while $m<0$ corresponds to deceleration
phase only. For tachyon coupled field, the expansion rate increases
by increasing $m$ which directly affects the contraction rate such
that blue region gets smaller. Also, the negative values of
parameter $m$ shows only decelerated expansion of the cosmic model.

\end{document}